# Residual nuclide formation in $^{206,207,208,nat}$Pb and $^{209}$Bi induced by 0.04-2.6 GeV Protons as well as in $^{56}$Fe induced by 0.3-2.6 GeV Protons

Yu. E. Titarenko[1], V. F. Batyaev[1,a], A. Yu. Titarenko[1], M. A. Butko[1], K. V. Pavlov[1], R. S. Tikhonov[1], S. N. Florya[1], S. G. Mashnik[2], A. V. Ignatyuk[3], W. Gudowski[4]

[1] Institute for Theoretical and Experimental Physics (ITEP), 117218 Moscow, Russia
[2] Los Alamos National Laboratory, Los Alamos, NM 87545, USA
[3] Institute for Physics and Power Engineering, 249020 Obninsk, Russia
[4] Royal Institute of Technology, S - 106 91 Stockholm, Sweden

**Abstract.** This work is aimed at experimental determination of independent and cumulative yields of radioactive residual product nuclei in the intermediate energy proton-irradiated thin targets made of highly isotopic enriched and natural lead ($^{206,207,208,nat}$Pb), bismuth ($^{209}$Bi), and highly isotopic enriched iron ($^{56}$Fe). 5972 independent and cumulative yields of radioactive residuals nuclei have been measured in 55 thin Pb and Bi targets irradiated by 0.04, 0.07, 0.10, 0.15, 0.25, 0.6, 0.8, 1.2, 1.4, 1.6, and 2.6 GeV protons. Besides, 219 yields have been measured in 0.3, 0.5, 0.75, 1.0, 1.5, and 2.6 GeV proton-irradiated Fe target. In both cases, the protons were extracted from the ITEP U-10 synchrotron. The measured data are compared with experimental results obtained elsewhere and with theoretical calculations by seven codes. The predictive power was found to be different for each of the codes tested, but was satisfactory on the whole in the case of spallation products. At the same time, none of the codes can describe well the product yields throughout the whole product mass range, and all codes must be further improved.

## 1 Introduction

Some of the current and pending nuclear projects (such as Accelerator-Driven Systems (ADS)) involve copious nuclear data. Since the requisite quantity of the data cannot be obtained experimentally, the projects have to use reliable computational codes after they being verified and tested by comparing with as many experimental data as possible.

In implementing the ISTC Project #2002 in 2002-2004, the ITEP team has measured the production cross sections of residual product nuclides in ~0.04, 0.07, 0.10, 0.15, 0.25, 0.40, 0.60, 0.80, 1.20, 1.60, and 2.60 GeV proton-irradiated thin $^{208,207,206}$Pb, $^{nat}$Pb, and $^{209}$Bi targets which are considered as ADS-target materials. This work presents part of the data thus obtained [1] and compares our data with calculation results by seven codes to find the predictive power of the models.

Moreover, in 2006, the ITEP team started implementing new ISTC Project #3266 to study ADS-structure materials such as $^{56}$Fe, $^{nat}$Ni, $^{nat}$Cr, $^{93}$Nb, $^{181}$Ta, and $^{nat}$W. The proton energies to be studied under the new Project are the same as in ISTC Project #2002. Besides, 0.3, 0.5, 0.75, 1.0, 1.5, and 2.6 GeV proton irradiations were made for the purpose of comparing with the GSI inverse kinematics data [2]. The respective data are presented in brief below.

## 2 Experiment

The 10.5-mm diameter (127-358 mg/cm$^2$) thin $^{208,207,206}$Pb, $^{209}$Bi, and $^{56}$Fe targets were exposed together with the Al monitors of the same diameter (127-254 mg/cm$^2$) to protons extracted from the external channel of the ITEP U-10 synchrotron. Use was made of the following nearly monoisotopic metal samples: $^{208}$Pb (0.87%$^{206}$Pb, 1.93%$^{207}$Pb, 97.2%$^{208}$Pb), $^{207}$Pb (0.03%$^{204}$Pb, 2.61%$^{206}$Pb, 88.3%$^{207}$Pb, 9.06% $^{208}$Pb), $^{206}$Pb (94.0%$^{206}$Pb, 4.04%$^{207}$Pb, 1.96% $^{208}$Pb), $^{nat}$Pb (1.4%$^{204}$Pb, 24.1%$^{206}$Pb, 22.1%$^{207}$Pb, 52.4% $^{208}$Pb), $^{209}$Bi>99.9%, and $^{56}$Fe (0.3% $^{54}$Fe, 99.5% $^{56}$Fe, 0.2% $^{57}$Fe, 0.05% $^{58}$Fe) and of the $^{27}$Al(p,x)$^{22}$Na monitor reaction. The proton fluence varied from $3.1 \cdot 10^{13}$ to $1.4 \cdot 10^{14}$ p/cm$^2$. The produced radioactive products were recorded by a GC-2518 type detector of a 1.8 keV resolution in the 1332 keV $^{60}$Co gamma-line. Within a single irradiation run, the samples were exposed to protons for about 3-6 months.

The spectra measured were processed by GENIE2000, with interactive fitting of each spectrum after its being processed automatically. The results of processing the spectra were used as input data for the SIGMA code used to determine the cross sections for production of the radionuclides found. The details of experimental techniques are described in [1,3].

Eventually, 5972 cross sections for production of residuals were determined in 55 Pb and Bi target measurements and 219 cross sections were determined in six Fe target experiments. The Final Report on ISTC Project#2002 [1] pre-sents numerical values and plots of the cross sections obtained in the Pb and Bi experiments. The measured data will be sent also to the EXFOR database.

---

[a] Presenting author, e-mail: vfb@itep.ru

## 3 Theoretical simulations

### 3.1 Pb&Bi data

The cross sections of the products measured were simulated by the following seven codes: **LAHET** (using both **Bertini** and **ISABEL** INC models) [6], CEM2k+GEM2 (**CEM03**), LAQGSM+GEM2 (**LAQGSM03**) in their 2003 and 2006 versions [7], **INCL4+ABLA** [8,9], **CASCADE** in its 2004 and previous versions [10], **CASCADO**, and **LAHETO** [11]. The latter two codes are the recent IPPE-devised modifications of the CASCADE and LAHET codes. In total, the simulated excitation functions together with the experimental data have been plotted in 884 figures [1]. Figs. 1 and 2 show part of our plots as examples. Quantitative comparisons were made using the mean squared deviation factor <F> described in [3].

To understand the degree to which various codes agree with experimental data in different nuclide production ranges, all the products were tentatively divided into four groups: spallation products (A>170), deep spallation products (140<A<170), fission products (30<A<140), and fragmentation products (A<30). Besides, the energy ranges were tentatively divided into three groups: low ($E_p$<0.1 GeV), medium (0.1 GeV<$E_p$<1.0 GeV), and high ($E_p$>1.0 GeV) energies. Table 1 presents the mean squared deviation factors <F> for each of the groups together with the average <F> values for all comparisons. To facilitate the analysis, three lowest <F> values are given in red and three highest <F> values in blue, within each of the comparison groups.

**A>170 (spallation products).** Most of the spallation products are predicted satisfactorily with <F> below 2.0 when averaged over all energies. In the near-target range of the products (A>200), the predictive power of the codes depends on proton energy. For instance, the CEM03 code predicts these products with <F>~1.5 at energies below 1 GeV, but underestimates them strongly (<F>~6.0) at energies above 1 GeV. On the contrary, the LAHET and LAQGSM03 codes predict the product with <F>~1.5-2.0 at energies above 1 GeV, but fail to predict them so well at lower energies (<F>~4.5). The INCL4+ABLA behavior is similar, namely, <F>~1.3-1.5 at $E_p$>0.1 GeV and <F>~6.0 at $E_p$<0.1 GeV. It should be noted that all codes give similar <F> values when averaged over all energies, the fact that makes it difficult to prefer any given code.

**140<A<170 (deep spallation products).** The predictive power of the codes deteriorates as the product nuclide mass decreases. It should be noted that the deterioration degree varies in different codes. For example, <F> increases up to about 1.9 for the BERTINI model, rises up to 2.3 for LAQGSM03, and increases up to 3.7 for INCL4+ABLA. The latter underestimates much the deep spallation products by overestimating their threshold energies. Judging by the <F> values, the CASCADE2004 code is much ahead of other codes (<F>=1.47 against <F> =1.81 for BERTINI) in this region.

**Table 1.** Mean squared deviation factors <F> for different energy ranges and reaction products with A>30.

| Code | Mass of a product (A) | | | Proton energy ($E_p$, GeV) | | | Total |
|---|---|---|---|---|---|---|---|
| | A>170 | 140<A<170 | 30<A<140 | $E_p$<0.1 | 0.1<$E_p$<1.0 | $E_p$>1.0 | |
| ISABEL | 1.81 | 1.81 | 2.87 | 4.88 | 2.13 | - | 2.16 |
| BERTINI | 1.75 | 1.93 | 2.75 | 4.26 | 2.06 | 1.97 | 2.10 |
| INCL4+ABLA | 1.90 | 3.74 | 2.22 | 4.63 | 2.18 | 2.13 | 2.25 |
| CASCADE | 1.77 | 2.01 | 6.93 | 4.93 | 3.93 | 2.44 | 3.25 |
| CASCADE-2004 | 1.93 | 1.47 | 5.54 | 6.54 | 3.23 | 2.42 | 2.94 |
| LAQGSM03 | 1.98 | 2.32 | 2.71 | 3.03 | 2.35 | 2.09 | 2.26 |
| CEM03 | 1.98 | 2.07 | 2.25 | 2.08 | 1.77 | 2.39 | 2.07 |
| CASCADO | 1.99 | 2.22 | 2.83 | 2.69 | 2.33 | 2.22 | 2.29 |
| LAHETO | 1.99 | 1.96 | 1.98 | 4.85 | 1.76 | - | 1.98 |

**Fission products**, which amount to about a third of all the measured and analyzed nuclides, are described by the codes worse compared with the spallation products. The INCL4+ABLA, CEM03, and LAHETO codes show the best predictive power for fission products with <F> ranging from 2.0 to 2.3. The INCL4+ABLA code shows an ambiguous agreement with the data. Namely, <F> remains to be high (up to 6.0) in the 120<A<140 range, where the fission products get overlapped with the deep spallation products. In the case of fission products with A<120, however, the agreement proves to be the best among all codes (<F> is from 1.5 to 2.0). The LAQGSM03 code shows a somewhat greater difference from experimental data (<F> reaches 4), but in the 80<A<110 range, <F> is about 2. The CASCADE code gives the worst result as regards convergence with fission products (<F> is up to ~20), which is much worse as compared to the rest of the codes.

**Fragmentation products** are much underestimated by all the codes tested. The calculations underestimate the measured yields of fragments by more than an order of magnitude. On the whole, the CEM03 and LAQGSM03 results are the nearest to experimental data.

### 3.2 $^{56}$Fe data

Apart from the codes used for the Pb&Bi data prediction, the MCNPX code was used also for simulation the $^{56}$Fe data. **MCNPX** (INCL, CEM2k, BERTINI, ISABEL models) [13] includes the basic versions of the incorporated codes with the parameters corresponding to the earlier code versions. Moreover, the recent codes CEM03.01 and LAQGSM03.01 were used together with their two supplementary versions G1 and S1: G1 uses the fission-like binary-decay model GEMINI instead of GEM2; S1 uses the multifragmentation model SMM of Botvina et al. (see details in [7]).

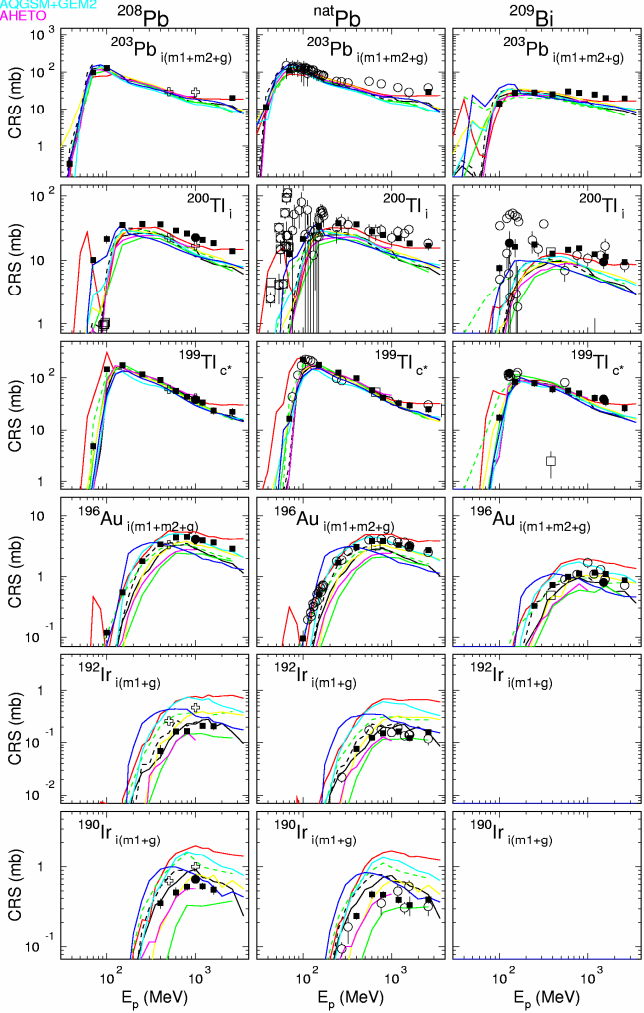
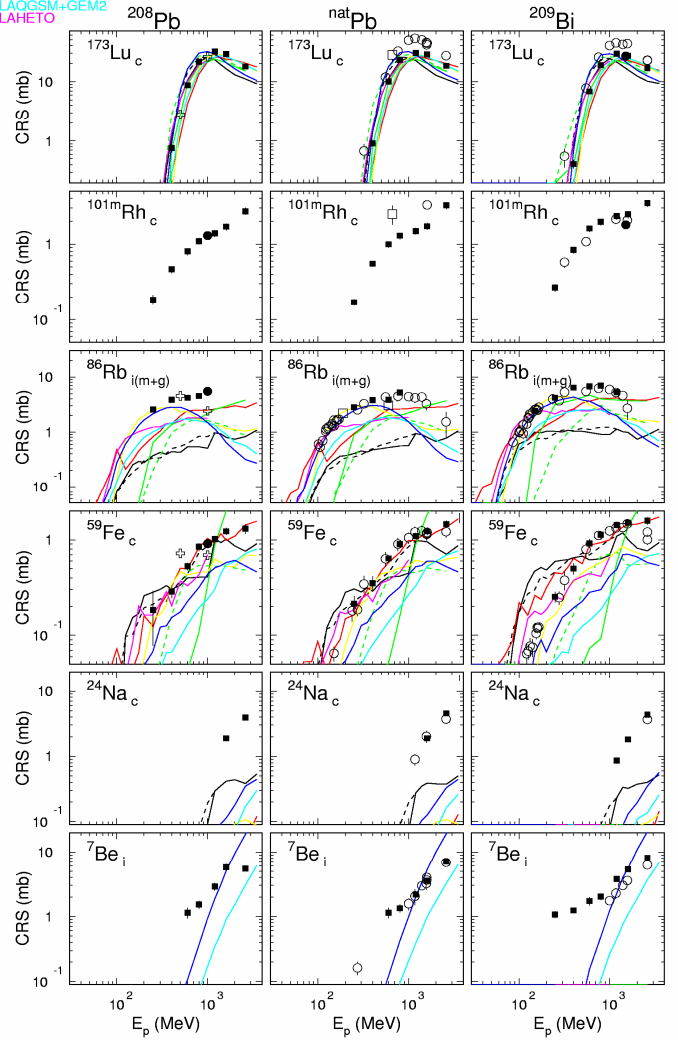

**Fig. 1.** The calculated and experimental excitation functions of $^{203}$Pb, $^{200}$Tl, $^{199}$Tl, $^{196}$Au, $^{192}$Ir, and $^{190}$Ir production in $^{208}$Pb (left), $^{nat}$Pb (center), and $^{209}$Bi (right) (■ show our data of this work, ● are our data measured earlier [3], ✢ are data measured at GSI via the inverse kinematics method [5], ○ are data measured at ZSR [4]. LAHET results are shown in black: ISABEL, as solid lines and BERTINI, as dashed lines; CEM03 results are in blue, INCL4+ABLA in red, CASCADE in green, LAQGSM03 in pale blue, LAHETO in purple, and CASCADO in yellow.

The results of comparison between nuclide production predictions by these models and $^{56}$Fe(p,x) experimental data are shown in Table 2. All the models give a relatively good description of nuclide production close to the target nucleus mass (A>30). In the mass range A<30, however, a high-quality description of the observed product nuclide yields is only given by the models that, apart from the conventional evaporation of complex particles, allow also evaporation of heavy clusters (the CEM and LAQGSM versions). So, our comparison with calculation results provides an impression that different reaction mechanisms dominate in each of the three mass ranges and, therefore, a qualitative representation of experimental data needs a more thorough simulation of each mechanism.

**Fig. 2.** The same as in Fig 2 for $^{173}$Lu, $^{101m}$Rh, $^{86}$Rb, $^{59}$Fe, $^{24}$Na, and $^{7}$Be.

## 4 Conclusion

In total, 6191 residual production cross section have been measured at ITEP in 61 experiments. The reliability of our measured data is proven via comparison with previous data measured elsewhere. The predictive powers of 14 models tested have been analyzed and proved to vary much. However, the predictive powers should be considered satisfactory for most of the nuclides in the spallation range. At the same time, none of the codes shows a good agreement with experimental data throughout the whole mass range of product nuclides, pointing thereby that all codes should be improved. On the whole, the predictive powers of all codes for fission products are worse compared with spallation products, and are even worse in the case of fragmentation products and at the spallation-fission interface. Therefore, further improvement of the evaporation/fission/fragmentation modes is a top priority task of researches in this field.

**Table 2.** Mean squared deviation factors $\langle F \rangle$ for $^{56}$Fe(p,x) reactions at different energies.

| Code/Model | Proton energy ($E_p$, GeV) | | | | | | Total |
|---|---|---|---|---|---|---|---|
| | 300 | 500 | 750 | 1000 | 1500 | 2600 | |
| MCNPX/INCL | 10.2 | 9.93 | 7.14 | 6.40 | 5.80 | 4.70 | **7.07** |
| MCNPX/CEM2k | 2.68 | 3.33 | 4.73 | 3.53 | 3.22 | 3.30 | **3.50** |
| MCNPX/BERTINI | 4.91 | 3.79 | 6.71 | 4.33 | 3.47 | 3.22 | **4.35** |
| MCNPX/ISABEL | 3.71 | 4.50 | 6.28 | 4.43 | 3.36 | 3.17 | **4.21** |
| LAHET/BERTINI | 4.41 | 4.14 | 3.26 | 4.86 | 4.01 | 3.57 | **4.02** |
| LAHET/ISABEL | 2.75 | 6.63 | 6.10 | 4.78 | 4.01 | 3.57 | **4.60** |
| CEM03.01 | 3.15 | 1.93 | 1.79 | 1.85 | 1.96 | 2.90 | **2.26** |
| CEM03.G1 | 2.61 | 2.58 | 2.56 | 2.30 | 2.19 | 2.81 | **2.51** |
| CEM03.S1 | 2.34 | 2.63 | 2.97 | 3.31 | 3.77 | 4.82 | **3.34** |
| LAQGSM03.01 | 5.04 | 2.75 | 2.24 | 2.16 | 2.15 | 3.32 | **2.87** |
| LAQGSM03.G1 | 3.85 | 2.41 | 2.57 | 2.57 | 2.65 | 3.72 | **2.95** |
| LAQGSM03.S1 | 2.95 | 2.62 | 2.87 | 2.89 | 2.94 | 4.02 | **3.06** |
| CASCADE-2004 | 2.82 | 2.78 | 4.44 | 4.41 | 5.27 | 5.79 | **4.30** |
| LAHETO | 3.90 | 5.44 | 4.46 | 6.18 | -- | -- | **5.02** |
| **Comparison with other experimental data** | | | | | | | |
| GSI-data | | 1.54 | 1.39 | 1.28 | 1.28 | 1.23 | - | **1.35** |

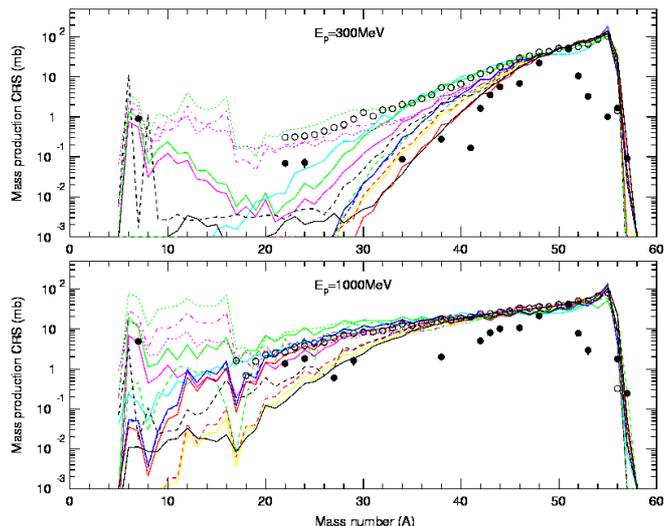

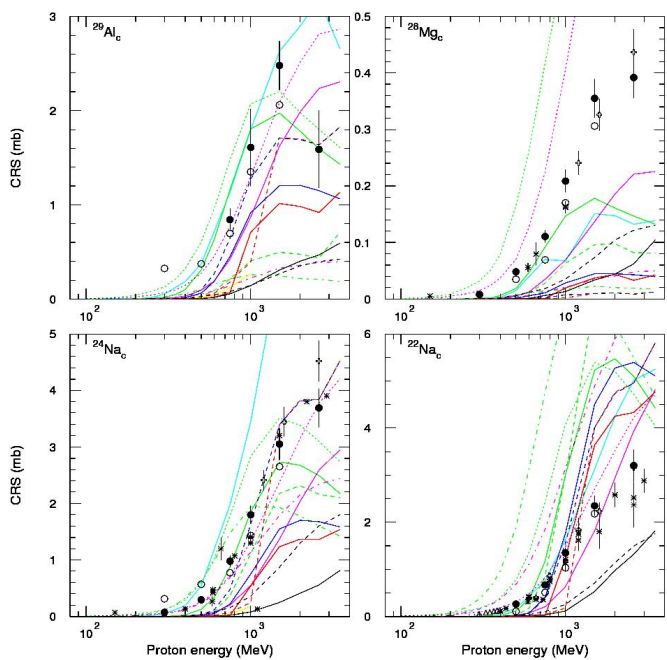

**Fig. 3.** Calculated and experimental excitation functions of $^{29}$Al, $^{28}$Mg, $^{24}$Na, $^{22}$Na production in $^{56}$Fe(p,x) reactions. ● shows our data, ○ are data measured at GSI via the inverse kinematics method [2], ✥ are data measured at ZSR [12], ✳ are other experimental data. MCNPX/INCL results are shown with black solid lines; MCNPX/BERTINI are in solid blue, MCNPX/ISABEL are in solid red; MCNPX/CEM2k are in dashed green; LAHET/BERTINI are in dashed blue; LAHET/ISABEL are in dashed red; LAQGSM03.01 are in solid magenta; LAQGSM03.G1 are in dotted magenta; LAQGSM03.S1 are in dashed-dotted magenta; CEM03.01 are in solid green; CEM03.G1 are in dotted green; CEM03.S1 are in dashed-dotted green; CASCADE-2004 are in pale blue; LAHETO are in yellow.

**Fig. 4.** Calculated and experimental dependences of residual nuclei yield mass distributions in the $^{56}$Fe(p,x) reaction at proton energies of 0.3 and 1.0 GeV. ● show our data, ○ show GSI data [2]. The line colors and types correspond to the code names as given in Fig. 3.

## 5 Acknowledgement

This work has been carried out under the EC-supported ISTC Projects#2002 and 3226. The work has also been supported by the Federal Atomic Energy Agency of Russia and, in part, by the U. S. Department of Energy at Los Alamos National Laboratory under Contract DE-AC52-06NA25396.